# Weaving the Future:
# Generative AI
# and the Reimagining of Fashion Design

Pierre-Marie Chauvin, Sorbonne Université
Hugo Caselles-Dupré, Obvious
Mathieu de Fayet, LVMH
Xavier Fresquet, Sorbonne Université
Angèle Merlin, Sorbonne Université
Benjamin Simmenauer, Institut Français de la Mode

**Introduction[1]**

Generative artificial intelligence has experienced a dramatic surge in the creative industries, particularly within the spheres of visual media and fashion over the past decade. Initially restricted to technical experimentation in research laboratories, these algorithms—capable of producing vivid imagery, synthesized text, and even plausible design concepts—have gradually permeated high-profile artistic and commercial platforms. The fashion sector, long heralded for its amalgamation of artistry, craft, and trend-setting innovation, is a fertile ground for these emergent technologies. By translating computational processes into visual or material outputs, generative AI challenges existing paradigms of creation and ignites debates on human authorship, copyright, and the ontological status of the finished piece[2]. This conflation of digital innovation with artisanal tradition thus positions fashion as both a bellwether of cultural evolution and an exemplar for broader industry shifts[3].

Within this confluence of art and technology, the paper interrogates two principal dimensions. First, it examines how AI-driven imagery and automation could contribute to the conception and realization of fashion collections. While early forays featured simplistic pattern generation or design augmentation, the sophisticated systems of today purport to simulate human creativity in ways that were once unimaginable. These processes—employed in moodboard creation, rapid prototyping, and virtual styling—invite reflection on the boundaries of technical efficiency versus artistic inspiration[4]. Second, the study investigates the reconfiguration of creativity, originality, and materiality wrought by AI interventions. For centuries, fashion has functioned as a barometer of aesthetic and cultural shifts; the rise of generative AI intensifies this dynamic by inserting computational logic into the creative process. Questions about authenticity, agency, and the very 'touch of the maker' thus resurface in dialogue with machine learning, prompting us to reconsider the essence of the fashion object[5].

The subsequent sections follow a logical progression to address the core issues at stake. The first explores conceptual foundations, tracing an intellectual genealogy of fashion's creative evolution and positioning AI within ongoing debates around authorial intention and the industrialization of craft. The second delves into the emergent ecosystem of AI-driven design, spotlighting changes in workflow, collaboration, and the reallocation of labor. Finally, the third section tackles broader socio-ethical ramifications, including legal disputes over ownership, environmental implications of high-volume computation, and the inclusivity or exclusivity of data-driven aesthetics[6].

---

[1] This paper draws upon the seminar "Tisser le futur : L'IA générative et le design de mode," held on 27 January 2025 at the Amphithéâtre Liard, Sorbonne Université. The event was recorded and subsequently transcribed; while generative AI tools were employed to assist in refining the materials, the final text was written and shaped by human contributors.

[2] Abbott, Ryan ; Rothman, Elizabeth, « Disrupting Creativity: Copyright Law in the Age of Generative Artificial Intelligence », *Florida Law Review*, Vol. 75, No. 6, December 2023, pp. 1141-1196.

[3] Talarico, Antonio, « Fashion's Next Generation: How Technology and Culture Are Combining », *The Interline*, March 12th 2024, 2025. (URL : https://www.theinterline.com/2024/03/12/fashions-next-generation-how-technology-and-culture-are-combining/)

[4] Zhang, Yanbo ; Chuanlan Liu, « Unlocking the Potential of Artificial Intelligence in Fashion Design and E-Commerce Applications: The Case of Midjourney », *Journal of Theoretical and Applied Electronic Commerce Research*, 18 mars 2024, no.1, pp. 654-670.

[5] Dennis, Caen A., « AI-generated fashion designs: Who or what owns the goods ? », *Fordham Intellectual Property, Media & Entertainment Law Journal,* 2020, Vol.30, No.2, pp. 593-625.

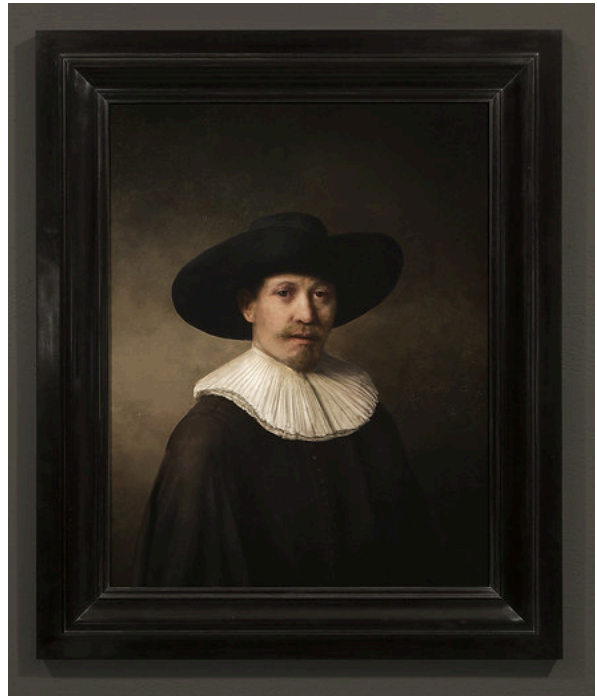

*Figure 1: The Next Rembrandt, an AI-generated painting created in 2016 using deep learning algorithms and facial recognition techniques. Developed by J. Walter Thompson Amsterdam, this project analyzed thousands of fragments from Rembrandt's existing works to generate a new portrait in the style of the Dutch master. The final artwork, printed in 3D to replicate the texture of oil paint, sparked discussions on AI's role in artistic creation and authenticity.*

---

[6] Doyle, Megan, « Why fashion should think carefully about using generative AI », *Vogue Business*, 6 mars 2025, 2025. (URL : https://www.voguebusiness.com/story/sustainability/why-fashion-should-think-carefully-about-using-generative-ai)

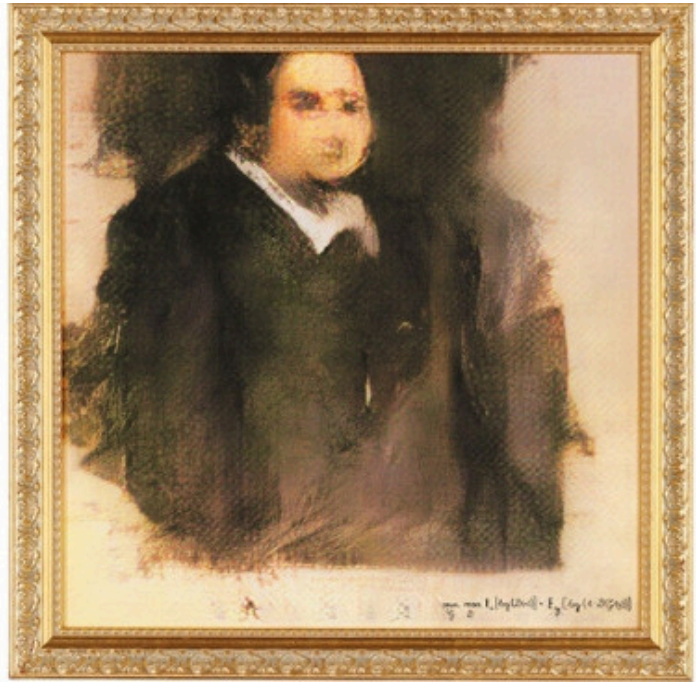

*Figure 2: Edmond de Belamy is a generative adversarial network (GAN) portrait painting constructed by Paris-based arts collective Obvious in 2018. Printed on canvas, the work belongs to a series of generative images called La Famille de Belamy. The print is known for being sold for US$432,500 during a Christie's's auction, marking the first artwork created with AI to be auctioned in a major auction house.*

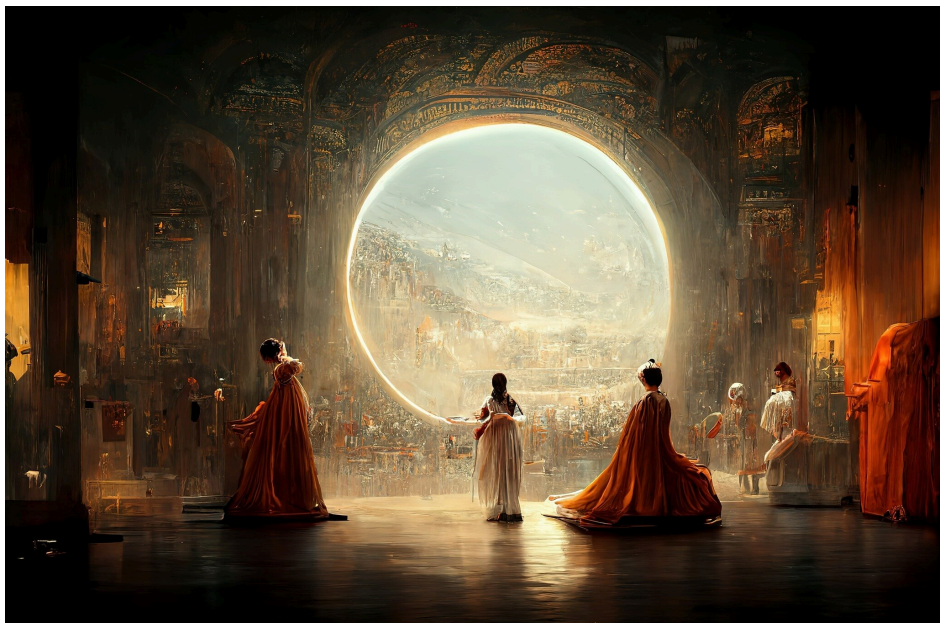

*Figure 3: Théâtre d'opéra spatial, an AI-generated artwork created using Midjourney by Jason Michael Allen in 2022. Winner of the fine arts competition at the Colorado State Fair, this image sparked debate on the role of AI in artistic creation and copyright issues, as the U.S. Copyright Office denied it legal protection.*

## I. Conceptual Foundations—Creativity and AI Technology in Fashion

Fashion has long been praised for blending artistic ingenuity, craftsmanship, and cultural commentary into tangible objects—garments worn on the body[7]. In the late nineteenth century, the introduction of industrial sewing machines ushered in a profound shift, as ateliers adapted to mechanized processes that promised both increased output and new forms of embellishment. This tension between craftsmanship and mass production foreshadowed later debates sparked by the digital revolution. Innovations such as computer-aided design (CAD) and digital pattern making further transformed the traditional workflow: though fashion's core often remained in the hands of designers, the conceptual tools at their disposal multiplied. These technological leaps did not replace the techniques of draping or tailoring; rather, they enhanced them, urging designers to fuse new methods with long-standing traditions.

The valorization of handcraft and symbolism stems from the belief that the maker's touch lends garments a distinctive aura. Whether in haute couture or refined *prêt-à-porter*, fashion depends on human judgment and culturally embedded codes. These appear in both the deft handling of fabric and in motifs that evoke historical or regional identity[8]. Yet, as mechanization advanced, questions about the source of creativity and authenticity in design grew more pressing. Was the seamstress or the machine responsible for the final line of stitches? This on-going debate, – centered on the tension between craftsmanship and technological intervention – has remained a central thread in fashion discourse, laying the groundwork for understanding the present-day implications of generative artificial intelligence.

Within this historical continuum, generative AI emerges as a formidable catalyst, reshaping fundamental assumptions about authorship, vision, and the aesthetic process[9]. Fundamentally, AI-generated design operates via neural networks—algorithms trained to detect patterns across vast image datasets and generate novel visual outputs. Systems such as Generative Adversarial Networks (GANs) and diffusion models absorb visual references by training on extensive image banks. Human input, through prompting, steers the AI toward specific stylistic references—ranging from eighteenth-century silhouettes to contemporary streetwear or abstract forms. The outcomes often open unexpected pathways beyond the reach of conventional design research.

Simultaneously, this collaborative framework prompts renewed inquiry into where creative intention ultimately resides. A perceptible shift emerges when AI-generated outputs display a coherence or originality not easily attributed to direct human input[10]. While designers insert prompts and set aesthetic parameters, the algorithm traverses expansive networks of reference—far exceeding human capacity for synthesis and recall. In this light, technology is no longer merely a facilitator but begins to appear as a co-creator. Yet, this perspective remains contested: for many practitioners, true creativity demands the tangible interplay of hand, eye, and intellect—an embodied alchemy that AI cannot reproduce.

One of the primary concerns surrounding AI's role in fashion lies in distinguishing newness from recombination: because generative models are trained primarily on existing data—fashion archives, runway imagery, and user-generated content—there is an abiding fear that these systems effectively "remix" prior styles rather than generate genuine innovations[11]. Critics argue that such recontextualization can result in designs that evoke a sense of *déjà vu*, blending familiar motifs from iconic collections with minor digital alterations. Proponents, by contrast, contend that all creative work involves reference and appropriation, and that AI's capacity to traverse billions of visual inputs may produce surprising outcomes.

Still, issues of bias and homogenization loom large if many designers adopt the same algorithmic tools. The aesthetic palette risks becoming constrained by the limitations of training data, leading to what some describe as a "data monoculture". Subtleties of local craftsmanship, vernacular styles, and subcultural fashions might be overshadowed by the standardized logic of algorithmic selection. In this context, the boundaries of originality become a site of contention, as stakeholders question whether AI-driven methods enlarge fashion's expressive possibilities or narrow them through hyper-digitized uniformity. The ongoing challenge lies in balancing AI's exploratory potential with a culturally attuned and critically sensitive articulation of the designer's human voice.

---

[7] Filieri, Juri ; Benelli, Elisabetta ; Filippi, Francesca, « Fashion design and art. Between Mutual Voracity and Disciplinary Self-Determination », *Fashion Highlight*, Juillet 2023, Vol. 1, pp. 88-95.
[8] Roche, Daniel, *The Culture of Clothing : Dress and Fashion in the Ancient Régime*, Cambridge, Cambridge University Press, 1997.
[9] Tsimitakis, Matthaios, « The AI Copyright Conundrum: Redefining Creativity in the Digital Age », *Creativesunite.com*, August 7th 2024, 2025. (URL : https://creativesunite.eu/article/the-ai-copyright-conundrum-redefining-creativity-in-the-digital-age )
[10] Borrelli-Persson, Laird, « Exactly What Is Copy, the First AI-Powered Fashion Magazine, Trying to Prove ? », *Vogue.com*, August 29th 2023. (URL : https://www.vogue.com/article/exactly-what-is-copy-the-first-ai-powered-fashion-magazine-trying-to-prove)
[11] Abbott, Ryan ; Rothman, Elizabeth, « Disrupting Creativity: Copyright Law in the Age of Generative Artificial Intelligence », *Florida Law Review*, Vol. 75, No. 6, December 2023, pp. 1141-1196.

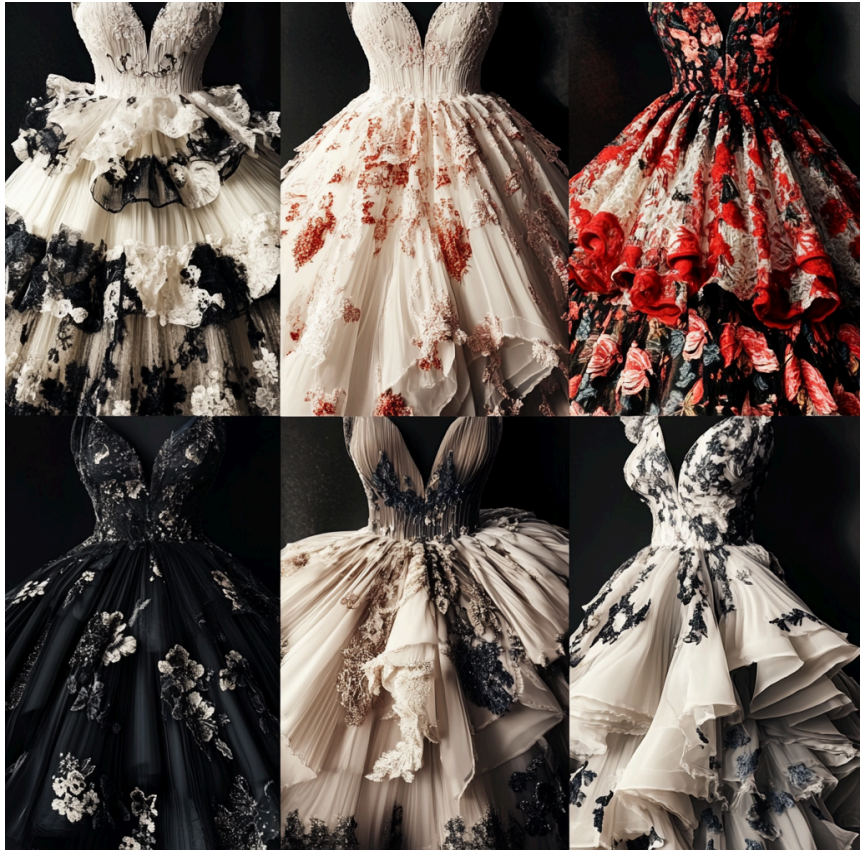

*Figure 4: AI-generated textile and silhouette designs inspired by Dior's haute couture aesthetic. Created using MidJourney, this visual sequence showcases fabric textures, embroidery, and drapery, reflecting the fusion of artisanal craftsmanship and generative AI in fashion design.*

**II. AI and creative renewal in fashion?**

Recent analyses have underscored the distinction between "generating" and "creating," particularly in the realm of AI-assisted fashion design. While AI, trained on extensive datasets, often recombines and rearranges existing visuals, genuine creation is typically understood as an "original" gesture intertwined with a specific artistic or philosophical vision. There might be a fine line between blank recombination and creative reinterpretation though: as many art historians and critics[12] have pointed out, creativity rarely operates outside of a context and most artworks, even the ones considered the most "original", derive from acquaintance with previous works and art traditions[13]. One should maybe move from the distinction between remix and true creativity to the finer-grained distinction between creative and non-creative *uses* of AI.

---

[12] Krauss, Rosalind E, *The originality of the avant-garde and other modernist myths*, MIT press, 1986.
[13] Levinson, Jerrold, « Refining art historically », *The Journal of Aesthetics and Art Criticism* 47.1, 1989, pp. 21-33.

Looking at processes and workflows in fashion studios, we can identify two key steps exemplifying creativity. One happens at an early stage of the development of collections and corresponds to the creative director's initial vision, that can be formulated and communicated to the other members of the creative team through different methods, with stories, themes, moodboards, sets of pictures, etc. The other has to do with the interactions within the company, among designers and the various technical experts (pattern makers, sample makers…), delivering prototypes and improving them until they reach the creative director's validation. These two steps, ideation and iteration, constitute the backbone of fashion's creative process within human organizations. It seems a pragmatic assumption to try to replicate this functioning as much as possible with the new possibilities of AI (see infra, section III).

Beyond the reproduction of human practices, AI can contribute in its own way, so to speak, to the creative process. For example, AI can give rise to unexpected "creative accidents": by occasionally introducing errors or surprising elements, it stimulates new directions in design thinking. Early experiments by fashion students and a few senior designers —where AI generated both silhouettes and patterns—illustrate the potential to discover fresh aesthetics. In this manner, AI ceases to be seen as merely a computational mechanism and becomes a partner in ideation, complementing rather than replacing the essential human aspect of craftsmanship.

Fashion involves a three-dimensional object created for the human body, necessitating a production chain that includes selecting materials, making prototypes, and performing manual adjustments. The "drape" of a garment, how it interacts with the wearer, and the tactile or visual sensations it provokes are crucial considerations. Since AI systems typically generate two-dimensional representations, moving from a virtual design to a tangible prototype remains technologically challenging. Nevertheless, emerging research is investigating the potential of simulating drape with computational models that can predict a fabric's behavior under various conditions. This approach has the added benefit of reducing the number of physical prototypes and lowering material waste. AI could also play a role in diversifying aesthetic standards by promoting inclusivity, although biases within training datasets may inadvertently reinforce narrow norms—particularly when certain silhouettes are overrepresented[14].

The use of AI in fashion raises multiple ethical and political questions. On the environmental front, the resource-intensive nature of data processing parallels the challenges of large-scale garment production. Advocates for a more sustainable approach view AI as a potential tool for better calibrating inventory and limiting unnecessary manufacturing by enabling more accurate demand forecasting and digital prototyping. Conversely, concerns persist that rapid, AI-driven design cycles might intensify consumerism by accelerating the introduction of new trends. Further inquiries center on data provenance and ownership, particularly regarding bias and transparency[15]. Another prominent issue is the place of human intention within AI-assisted creativity: while some worry about standardization and a possible loss of authenticity, others highlight the expanded creative horizons AI can unlock. Ultimately, AI functions as a catalyst for re-examining aesthetic values and the role of the designer's hand, prompting the fashion industry to reconsider its longstanding relationship with art, materials, and innovative processes.

**III. Reimagining the Fashion Design Process**

The integration of AI into contemporary fashion practices manifests most visibly at two critical stages of the design process: ideation and prototyping. During the ideation and moodboarding phase, designers employ AI-based image generators to unanticipated color palettes, motifs, and silhouette variations[16]. In practice, this can involve formulating textual "prompts" referencing a historical eras or conceptual themes —such as "Rococo streetwear" or "biomorphic knit textures"—, as well as using images of various kinds as style inspirations, from which the algorithm translates into a spectrum of novel visuals[17]. These outputs, though driven by a computational model, may resonate with or disrupt a designer's initial vision, expanding possibilities beyond the conventional array of references found in books or online image libraries. Notably, some studios curated "prompt banks" alongside the more traditional collage boards, suggesting a formal recognition of AI-driven exploration as integral to the creative process[18].

---

[14] D'Ignazio, Catherine, Klein, Lauren F., *Data Feminism*, The MIT Press, Cambridge, Massachusetts, 2020.
[15] Hayles, Katherine, *Unthought: The Power of the Cognitive Nonconscious*, The University of Chicago Press, Chicago, 2017.
[16] Anbouba, Margaux, « Collina Strada's AI-Inspired Clothing Was Balanced Out With "Snatural," Organic Beauty », *Vogue.com*, September 9th 2023. (URL : https://www.vogue.com/slideshow/collina-strada-ss24-beauty)

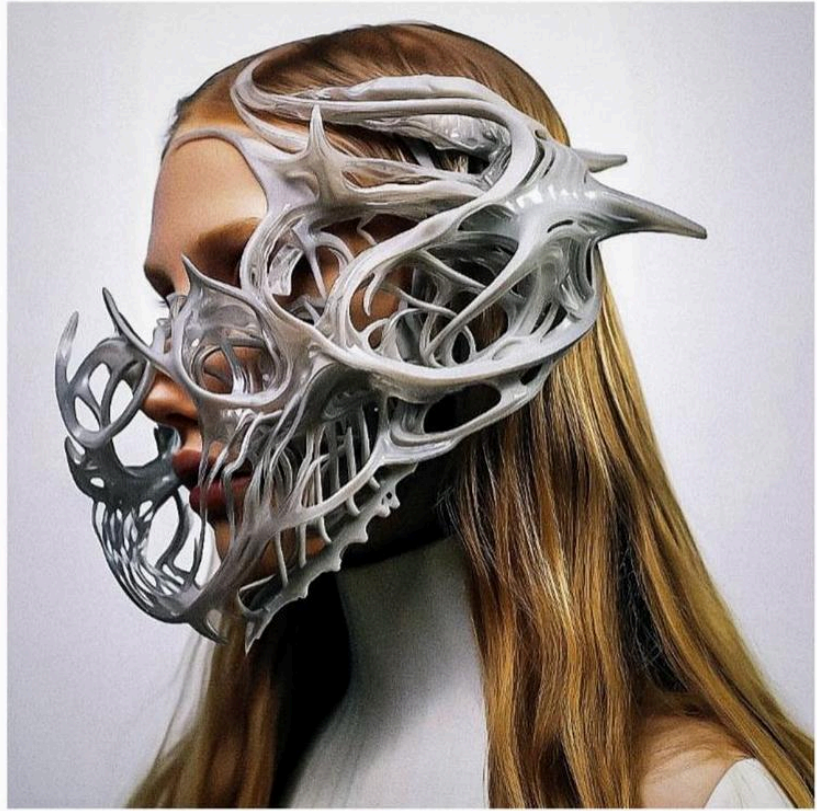

*Figure 5: AI-generated wearable by Jean Ronel Aldric Tingbo, featured in [Prompt Magazine](), Book 6. The artist's approach is described as an exploration of « Wearables born from the crossing of fashion, fiction and futurism with a distinct biopunk / biomorphic aesthetic »*

Following the initial burst, rapid prototyping and simulation further accelerate the workflow. AI-assisted virtual fitting tools can approximate the drape of various fabrics, adjusting garment proportions in real time on digital avatars[19]. This approach stands in contrast to traditional sample-making, where each physical iteration entails material waste and prolonged lead times. By substituting or at least significantly reducing the number of initial toile prototypes, studios shorten time-to-market while gleaning data-driven insights into silhouette viability. Although these simulations cannot fully replicate the tactile nuance of real textiles, they effectively eliminate early-stage design errors and allow designers to focus on refinement. However, this technical efficiency often comes at the cost of creativity. Specialized applications achieve such speed by restricting the possible parameters of silhouette design, favoring standardized forms over true experimentation. As a result, while these tools align well with the demands of fast fashion and commercially oriented production, they may prove less suitable for designers seeking to push the boundaries of form and materiality.

---

[17] Dueno, Carlos ; Lopez-Figueroa, Johnny, « Fashion and A.I.: Determining How the Creative Process for Independent Designers Is Shifting in 2023 », *Journal of Student Research*, vol. 13, no. 2, May 2024.
[18] Bain, Marc, « Can AI Carry a Designer's Legacy ? », January 30th 2024. (URL : https://www.businessoffashion.com/articles/technology/can-ai-carry-on-a-designers-legacy/)
[19] Lee, Adriana, « The Rise of the Digital Fashion Model », *WWD.com*, 16 septembre 2022, 2025. (URL : https://wwd.com/business-news/technology/rise-of-digital-fashion-model-photogenics-1235335875/)

Despite its apparent efficiency, the emergent design methodology brings forward new questions regarding the balance between human vision and machine suggestion[20]. Designers often navigate a delicate negotiation: they submit wisely-chosen prompts to AI in order to guide the overall aesthetic. As Carl-Axel Wahlström, creator of *Copy Magazine*, notes in an interview with Laird Borrelli-Persson, crafting effective prompts is far from straightforward: « Creating images in AI really [requires that you] be very inventive with words. Often when I type words that are very clear to me the AI totally misunderstands me because the training models and the metadata that we're putting into our images gets confused »[21].

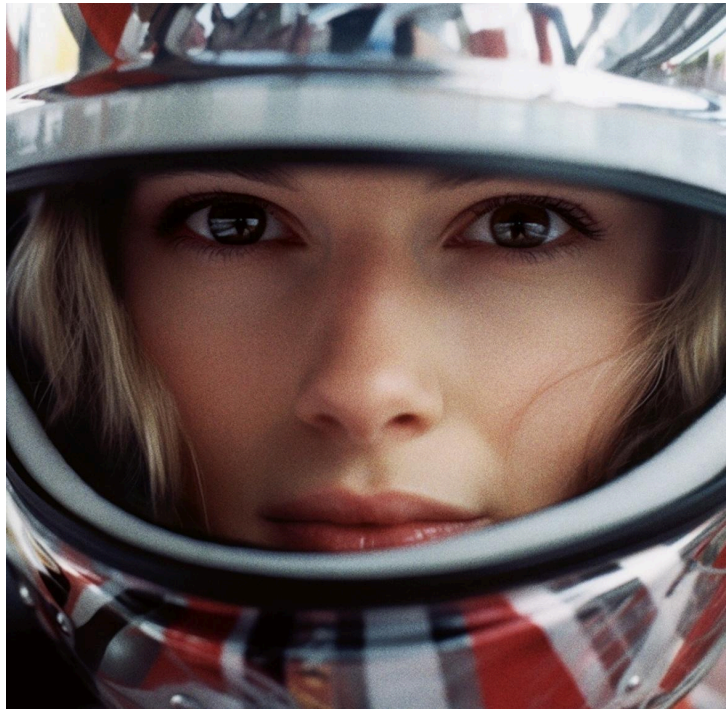

*Figure 6: AI-generated cover for Copy Magazine.*

The interplay between human input and algorithmic output can lead to results that surpass the initial intention, particularly when unexpected variations arise. When a designer discerns potential in a system-suggested texture, color fusion, or pattern synergy, a genuine co-creative relationship begins to take shape. Carl-Axel Wahlström expands on this idea, highlighting the generative potential of such misunderstandings: « It's also interesting because you discover things that you might not have thought about because we are also very limited in our fantasies and ideas of what to create. Many times when the AI misunderstands you, it comes up with something more interesting than what you could have realized. »[22].

---

[20] « "My concern with AI in the fashion world – and the broader creative world – is that it isn't collaborative or spontaneous," explains writer and culture critic, Charlie Squire. "A computer programe can design something interesting, something 'new,' but that thing lacks the conversational process of contextualisation that art has. And without that context, I think our clothes (and thus ourselves) will feel increasingly detached and unfulfilled."» Quoted in Montgomery, Joy, « I asked ChatGPT To Create My Outfits For A Week – This Is What Happened », *Vogue.com*, January 26th 2024, 2025. (URL : https://www.vogue.com/article/i-asked-chatgpt-to-create-my-outfits-for-a-week-this-is-what-happened)
[21] Borrelli-Persson, Laird, « Exactly What Is Copy, the First AI-Powered Fashion Magazine, Trying to Prove ? », *Vogue.com*, August 29th 2023. (URL : https://www.vogue.com/article/exactly-what-is-copy-the-first-ai-powered-fashion-magazine-trying-to-prove)
[22] Borrelli-Persson, Laird, « Exactly What Is Copy, the First AI-Powered Fashion Magazine, Trying to Prove ? », *Vogue.com*, August 29th 2023. (URL : https://www.vogue.com/article/exactly-what-is-copy-the-first-ai-powered-fashion-magazine-trying-to-prove)

Therefore, AI may act as a catalyst for conceptual leaps and "accidental" discoveries[23]. Algorithms that conflate multiple references may generate silhouettes impossible to envision through linear, human-led brainstorming. This capacity to unearth unpredicted directions exemplifies the generative potential of machine-led exploration, revitalizing design when inspiration falters. A striking example of this is Walter van Beirendonck's Spring/Summer 2024 Menswear collection, which openly embraced AI as a source of inspiration[24]. The collection seemed to echo the visual logic of machine-generated imagery—disruptive, playful, and provocatively unrefined. By channeling the unpredictable outputs of AI, Van Beirendonck underscored the value of embracing algorithmic "misfires" as fertile ground for innovation. At the same time, detractors caution that over-reliance on accidents may erode the rigor and intentionality traditionally celebrated in couture disciplines.

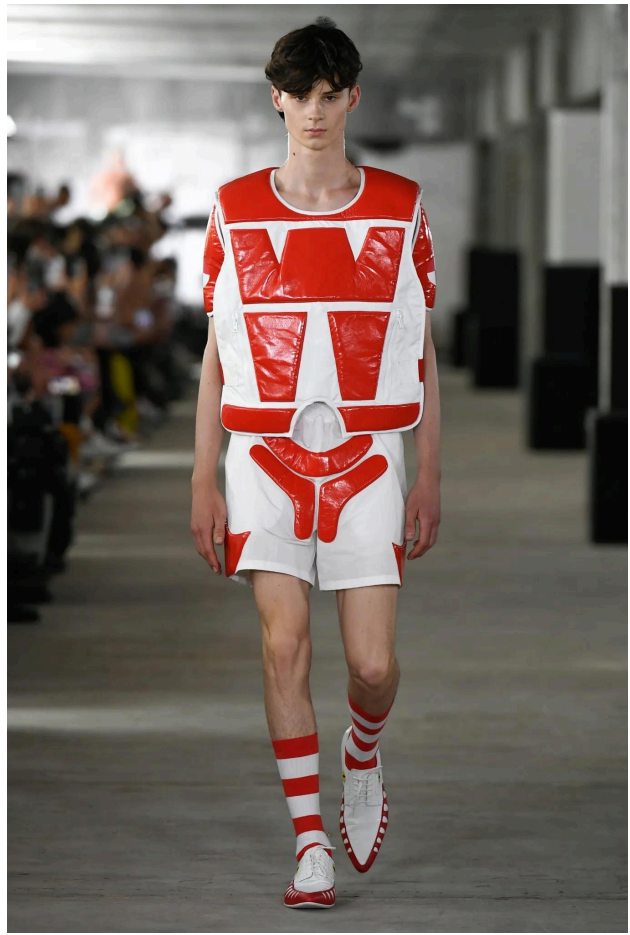

*Figure 7: Walter van Beirendonck, Spring / Summer 2024 Menswear collection inspired by AI.*

While AI excels at producing compelling digital imagery, the transition from two-dimensional concepts to wearable garments presents significant challenges[25]. Patternmaking, for instance, demands precise manipulations and seam placements that hinge on the interplay of fabric weight, grain, and drape. Even advanced simulation software cannot wholly capture how a particular silk might respond to the body's kinetic movements, nor how certain synthetic blends may stretch or shrink under tension. Consequently, the essential skills of tailors, patternmakers, and textile specialists retain their centrality in the production chain, ensuring that garments meet both aesthetic and functional standards.

---

[23] Borrelli-Persson, Laird, « Exactly What Is Copy, the First AI-Powered Fashion Magazine, Trying to Prove ? », *Vogue.com*, August 29th 2023. (URL : https://www.vogue.com/article/exactly-what-is-copy-the-first-ai-powered-fashion-magazine-trying-to-prove)
[24] Odunayo, Ojo, « Walter Van Beirendonck Spring 2024 Menswear », *Vogue.com*, June 21st 2023. (URL : https://www.vogue.com/fashion-shows/spring-2024-menswear/walter-van-beirendonck)
[25] Rajvanshi, Ashta, « How AI Could Transform Fast Fashion for the Better – And Worse », *Time.com*, September 20th 2024, 2025. (URL : https://time.com/7022660/shein-ai-fast-fashion/)

Ultimately, artisanal expertise remains indispensable—especially at the upper echelons of design, where tactile intuition is seen as integral to fashion's identity[26]. The tactile evaluation of a toile's curvature continues to surpass what current digital tools can approximate. At the same time, platforms such as [Refabric](#) or [Clo3D](#) demonstrate how AI can significantly accelerate the prototyping phase by translating rough sketches or moodboard cues into sophisticated 3D garment visualizations. By simulating fabric drape, movement, and structural behavior, these platforms could facilitate rapid iteration and test bold ideas before committing to physical materials. Thus, while AI-assisted design can streamline early prototyping, it is grounded in a centuries-old tradition of working intimately with materials. Rather than rendering craft obsolete, these technologies invite a redefinition of each step in the process: human artisans refine the details and imbue sensorial depth, while AI tools provide efficiency and a seemingly limitless reservoir of fresh concepts. This evolving partnership suggests that the future of fashion may lie in striking an equilibrium where creative autonomy, computational exploration, and material expertise converge.

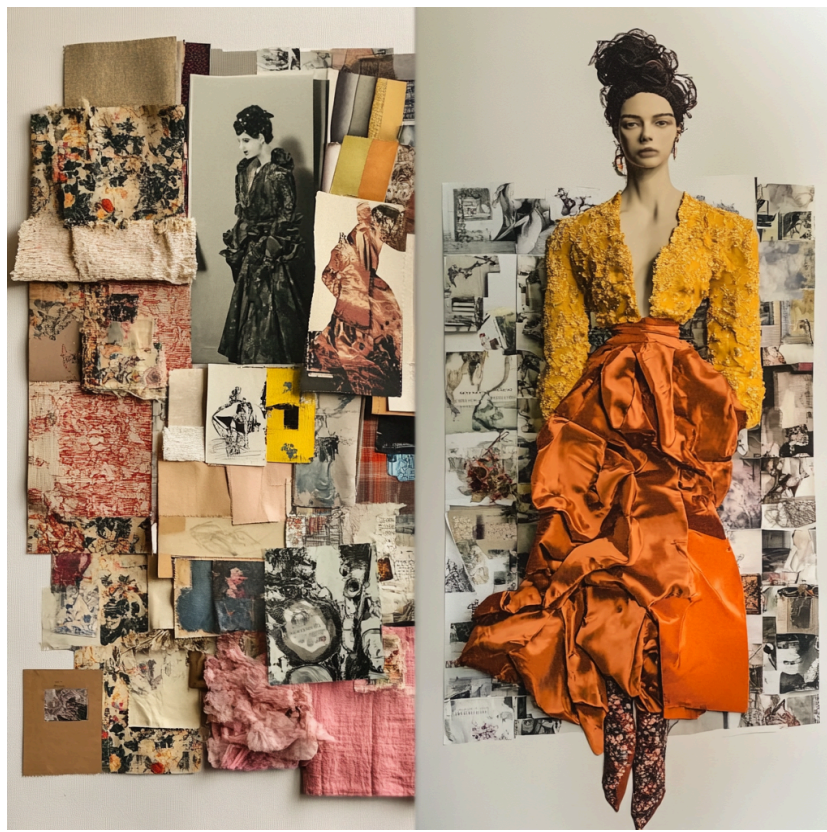

*Figure 8: A side-by-side comparison of a traditional collage-based moodboard and an AI-generated collage. This visual juxtaposition highlights the interplay of color theory, historical references, and thematic inspirations, showcasing the contrast between handcrafted design processes and AI-driven creativity in fashion and art.*

---

[26] Ramos, Leo ; Rivas-Echeverría, Francklin, Pérez, Anna Gabriela et al., « Artificial intelligence and sustainability in the fashion industry: a review from 2010 to 2022 », *SN Applied Science*, Vol. 5, No.387, 2023.

## IV. Debate Over AI-Generated Fashion

In the realm of fashion, new technologies often provoke sharp responses—ranging from enthusiastic acceptance to profound skepticism[27]. Generative AI, with its capacity to propose unconventional silhouettes or hybridize cultural motifs, has undoubtedly captured the public imagination. This fascination manifests in media coverage celebrating its futuristic allure, and in dedicated exhibitions focused on digital couture. Yet, concurrent voices express distrust regarding authenticity, contending that machine-derived concepts may lack the conceptual depth or emotional resonance traditionally attached to a designer's personal touch. Some critics posit that algorithmic output, reliant on broad datasets, risks flattening cultural nuance and rendering creativity overly formulaic.

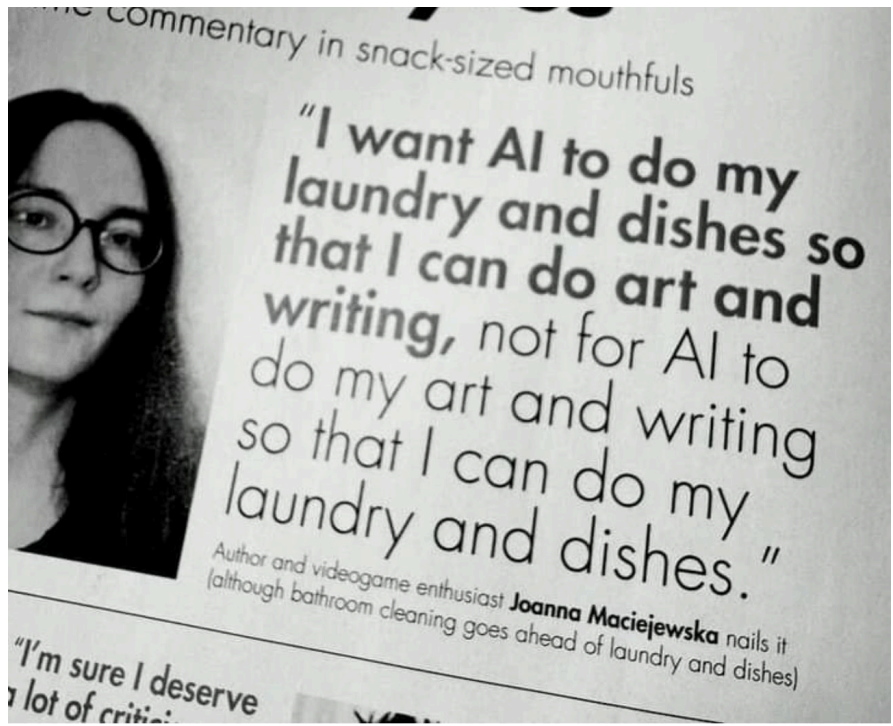

*Figure 9: Joanna Maciejewska's viral quote : « I want AI to do my laundry and dishes so that I can do art and writing, not for AI to do my art and writing so that I can do my laundry and dishes ».*

Beyond authenticity, AI's influence on trend cycles raises further concern. AI-enabled tools expedite the design process, potentially yielding a more rapid turnover of micro-trends that can intensify "fast fashion" phenomena[28]. With AI swiftly scanning consumer data and social media cues, brands might intensify production and obsolescence cycles to meet perceived market desires. Conversely, certain commentators suggest that AI could foster sustainable practices by eliminating unnecessary prototypes or facilitating on-demand manufacturing. From minimizing excess inventory to personalizing products for niche markets, these technological interventions might help reconcile efficiency with responsible resource management. Whether AI becomes a catalyst for reckless consumption or a vehicle for conscious design ultimately depends on the intentionality of both fashion houses and their consumers.

The rise of generative AI is also reshaping labor dynamics, as some skill sets become redundant and others rise to prominence[29]. Tasks once performed by patternmakers or illustrators—often

---

[27] TREFF, Michael, « The Fashion Industry Treat Tech Like a Seasonal Trend », *On_discourse.com*, 2025. (URL : https://ondiscourse.com/the-fashion-industry-treats-tech-like-a-seasonal-trend/
[28] Rajvanshi, Ashta, « How AI Could Transform Fast Fashion for the Better – And Worse », *Time.com*, September 20th 2024, 2025. (URL : https://time.com/7022660/shein-ai-fast-fashion/)
[29] Rajvanshi, Ashta, « How AI Could Transform Fast Fashion for the Better – And Worse », *Time.com*, September 20th 2024, 2025. (URL : https://time.com/7022660/shein-ai-fast-fashion/)

time-intensive—are increasingly automated, shifting these roles from hands-on production to creative oversight and quality control. In parallel, new hybrid roles emerge—positions that merge data analysis, programming proficiency, and creative flair. The ability to code an AI model or to refine algorithmic outputs becomes an increasingly valuable asset, particularly in mid-to-large fashion houses seeking to integrate AI-driven efficiencies into their daily operations. This transition underscores the importance of cross-disciplinary educational programs. By blending coursework in fashion design, data science, and ethics, emerging curricula aim to cultivate designers who navigate both computational technologies and aesthetic traditions with equal fluency[30].

Dutch designer Iris van Herpen offers a nuanced theoretical lens for rethinking the evolving relationship between craftsmanship and technological innovation. While her practice does not explicitly engage with artificial intelligence, her concept of "craftolution"—a term denoting the synthesis of artisanal techniques and technological advancement—proposes an epistemological parity between manual and digital processes. As van Herpen articulates, « We combine the organic and the synthetic to blur the line between nature, craftsmanship and technology »[31]. Her perspective foregrounds a reconciliatory approach—positioning technological mediation not as a rupture from tradition, but as a continuation of it. In this view, digital tools are absorbed into the lineage of craftsmanship, enabling new forms of expression while preserving the ontological depth of hand-based making.

---

[30] Allaire, Christian, « Ben Barry, the New Dean at Parson's, Is All In on Inclusive Fashion », *Vogue.com*, December 11th 2024. (URL : https://www.vogue.com/article/ben-barry-parsons-school-of-design-inclusivity-diversity).
[31] Quote by Iris van Herpen. See in Obrist, Hans-Ulrich, Herpen (van), Iris, « Iris van Herpen », interview published in Pitiot, Cloé, Iris van Herpen. Sculpting the Senses, Musée des arts décoratifs, Paris, 2023, pp.24-43.

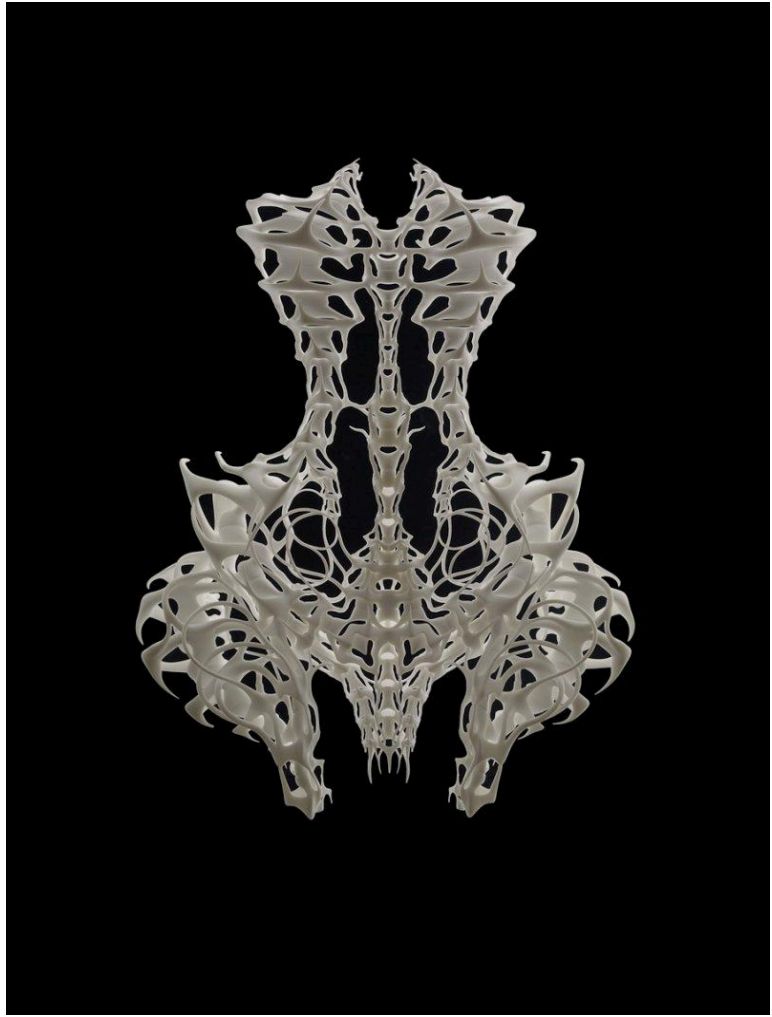

*Figure 10: Iris van Herpen "Skeleton" set, Fall / Winter 2011-2, "Capriole" collection, Haute couture. 3D-printed white polyamide (plastic) (SLS); leather; acrylic (plastic), The Metropolitan Museum of Art, New York. Asc. n°2012.560a-d*

While such programs empower a new generation of creative technologists, they also bring potential pitfalls. Fears of deskilling persist, particularly if certain artisanal competencies—like hand illustration or pattern grading—cease to be transmitted from master to apprentice. Some ateliers safeguard these traditions through dedicated mentorship, recognizing their aesthetic and cultural value despite the allure of digital replacements. is to balance innovation with preservation—ensuring that the industry's historical techniques and narratives endure amid influx of algorithmic tools. At the intersection of law and fashion, questions of data ownership and copyright loom large in discussions surrounding AI-generated design. Since these algorithms learn from massive repositories of existing imagery—ranging from archived runway photos to streetwear snapshots—the question arises: how do we differentiate inspiration from infringement?[32] The risk of plagiarism intensifies if an AI model inadvertently reproduces near-identical motifs from the copyrighted materials they were trained on. In response, some technology firms advocate for "transparent training logs" or licensing agreements that clarify usage rights—though standardized regulation remains nascent within the fashion sector.

Equally pressing is the potential for bias in training datasets, which may inadvertently perpetuate cultural stereotypes or exclude minority aesthetics altogether[33]. Designers and brands, working alongside AI developers, bear a responsibility to audit these datasets, ensuring that generative outputs neither appropriate marginalized communities without credit nor reinforce harmful clichés. Conversations around cultural appropriation thus intersect with concerns about algorithmic bias: datasets lacking diversity can yield designs that replicate inequities beneath a veneer of innovation. Meanwhile, the environmental footprint of large-scale computation poses another ethical dimension[34]. While AI-based prototyping could reduce physical samples, the energy consumption tied to server farms and high-performance hardware could offset potential sustainability gains. Responsible adoption, therefore, demands a holistic approach—aligning computational tools with eco-conscious production strategies. By treating data ethics, inclusivity, and environmental stewardship as core design principles, the fashion industry can harness AI's potential while mitigating its unintended consequences.

---

[32] Bain, Marc, « Can AI Carry a Designer's Legacy ? », January 30th 2024. (URL : https://www.businessoffashion.com/articles/technology/can-ai-carry-on-a-designers-legacy/)

[33] D'Vaz-Sterling, Aasia, « Fashion Has Been Tarnished By Inequality. Can Generative AI Help Deliver A More Equitable Future ? », *Theinterline.com*, July 1st 2024. (URL : https://www.theinterline.com/2024/07/01/fashion-has-been-tarnished-by-inequality-can-generative-ai-help-deliver-a-more-equitable-future/)

**Conclusion**

We've discussed how, when thoughtfully integrated, generative AI can enrich fashion design—extending aesthetic frontiers and introducing new forms of operational efficiency[35]. From prompting novel silhouettes in the early ideation stages to aiding with rapid prototyping, AI-equipped workflows promise to streamline production and spark serendipitous breakthroughs. Nonetheless, these same tools mandate consistent ethical and cultural checks. The potential for algorithmic bias, inadvertent appropriation, and environmental strain compels designers, technologists, and brand executives to remain vigilant. By acknowledging the delicate balance between authentic craftsmanship and computational exploration, the industry preserves an identity shaped by personal touch and storied heritage, even as it welcomes new forms of machine-assisted originality.
Looking ahead, a clear call emerges for cross-sector collaboration: designers shaping aesthetic discourse; technologists refining algorithms; academics offering critique and context; and policymakers developing frameworks that promote inclusivity, sustainability, and creative freedom[36]. Such alliances may refine standard practices, including robust data management, equitable design crediting, and fair labor transitions.

Moreover, emerging frontiers invite further research: multi-sensory experimentation that bridges visual output with tactile or olfactory experiences; advanced robotics capable of integrating real-time AI feedback into garment construction; and personalized digital fittings that adapt clothing to individual body metrics. Each of these avenues reveals an expanding horizon where machine learning coexists with—rather than replaces—human ingenuity[37].

Though AI has the capacity to transform core processes—digitizing mood boards, generating fresh silhouettes, or compressing design timelines—it cannot negate the deeply anthropological dimensions at the heart of fashion. Identity, storytelling, and emotional resonance remain central to how garments signify cultural belonging and personal expression. The interplay of stitchwork, material manipulation, and lived tradition cannot—nor should it—be fully automated. By envisioning a future in which craft and technology interact, practitioners create a vibrant ecosystem wherein each is enriched by the other. As AI continues to shape the future of fashion, can the industry weave an inclusive, innovative fabric—one that honors its heritage while embracing the creative possibilities still unfolding?

---

[34] « Moreover, the fast fashion model has accelerated the production and disposal of clothing, resulting in a surge in waste generation and unsustainable resource consumption. As a countermeasure to all these problems, the fash‑ion industry is increasingly trying to find solutions and tools that will enable it to achieve sustainability goals. As a result, the fashion industry is increasingly turning to AI to help improve sustainability » In Ramos, Leo ; Rivas-Echeverría, Francklin, Pérez, Anna Gabriela et al., « Artificial intelligence and sustainability in the fashion industry: a review from 2010 to 2022 », *SN Applied Science*, Vol. 5, No.387, 2023.
[35] Min, Yoo-Won Olivia, and B. Ellie Jin, « Here's How the Fashion Industry Is Using AI », *Wilson College News*, June 3rd, 2024. (URL : https://www.vogue.com/article/ben-barry-parsons-school-of-design-inclusivity-diversity)
[36] World Economic Forum, « Technology Policy: Responsible Design for a Flourishing World », October 2024. (URL : https://www3.weforum.org/docs/WEF_Technology_Policy_Responsible_Design_Flourishing_World_2024.pdf)
[37] Première Vision, « Creativity in the Age of AI: From Concept to Realisation », *Première Vision Magazine*, February 24th 2025. (URL : https://www.premierevision.com/en/articles/creativity-in-the-age-of-ai-from-concept-to-realisation/)